\input{aipcheck}

\documentclass[
    ,final            
  ]
  {aipproc}

\layoutstyle{8x11single}

\usepackage{graphics}
\usepackage{epsfig}
\usepackage{bm}
\usepackage{epic,eepic}
\usepackage{epsfig}
\usepackage{pifont}
\usepackage{subfigure}
\usepackage{caption}
\usepackage{graphicx}
\usepackage{amsmath,amssymb,amsfonts,mathrsfs}
\usepackage{url} 

\usepackage[skip=5pt,font={bf,sc,small},singlelinecheck=off]{caption}

\newcommand{\be}{\begin{equation}}
\newcommand{\ee}{\end{equation}}

\usepackage{float}

\usepackage{mathptmx}
\usepackage{latexsym}



\begin{document}

\title{Viscoelastic Multicomponent Fluids in confined Flow-Focusing Devices}


\author{A. Gupta}{
  address={Department of Physics and INFN, University of ``Tor Vergata'', Via della Ricerca Scientifica 1, 00133 Rome, Italy}
}

\author{M. Sbragaglia}{
  address={Department of Physics and INFN, University of ``Tor Vergata'', Via della Ricerca Scientifica 1, 00133 Rome, Italy}
}

\author{E. Foard}{
  address={Department of Physics and INFN, University of ``Tor Vergata'', Via della Ricerca Scientifica 1, 00133 Rome, Italy}
}

\author{F. Bonaccorso}{
  address={Department of Physics and INFN, University of ``Tor Vergata'', Via della Ricerca Scientifica 1, 00133 Rome, Italy}
}

\begin{abstract}
The effects of elasticity on the break-up of liquid threads in microfluidic cross-junctions is investigated using numerical simulations based on the "lattice Boltzmann models" (LBM). Working at small Capillary numbers, we investigate the effects of non-Newtonian phases in the transition from droplet formation at the cross-junction (DCJ) and droplet formation downstream of the cross-junction (DC) (Liu \& Zhang, {\it Phys. Fluids.} {\bf 23}, 082101 (2011)). Viscoelasticity is found to influence the break-up point of the threads, which moves closer to the cross-junction and stabilizes. This is attributed to an increase of the polymer feedback stress forming in the corner flows, where the side channels of the device meet the main channel.
\end{abstract}

\maketitle


\section{Introduction}

The progressive breakup of a fluid thread into a number of small drops is a rich physical phenomenon~\cite{Eggers97}. It is also an important dynamical process that impacts many applications. In particular, droplet-based microfluidic devices have gained a considerable deal of attention  due to their importance in studies that require high throughput control over droplet size~\cite{Teh08,Baroud10,Seeman12,Glawdeletal}.  Common droplet generator designs used in these devices are T-shaped geometries~\cite{Demenech07} and flow-focusing devices~\cite{LiuZhang09,LiuZhang11,Garstecki13}. In T-shaped geometries, a dispersed (d) phase is injected perpendicularly into the main channel containing a continuous  (c) phase. Forces are created by the cross-flowing continuous phase which periodically break off droplets; the flow-focusing geometry, instead, creates drops by focusing a fluid stream into another co-flowing immiscible fluid~\cite{Gordillo,Link04,Anna,LiuZhang09,LiuZhang11}. The operational regime of these devices is primarily characterized by the Capillary number $Ca$, which quantifies the importance of the viscous forces with respect to the surface tension forces at the non-ideal interface and the drop size and its distribution are dictated by the relative flow rates of the two fluids. With few exceptions~\cite{Arratia08,Steinhaus,Husny06}, previous research has been mainly restricted to Newtonian fluids. The processing of biological fluids with high molecular weight macromolecules, inevitably results in considering a non-Newtonian (NN) behaviour. Consistently, the use of viscoelastic liquids in flow-focusing devices~\cite{Arratia08,Arratia09,Steinhaus} or T-junction geometries~\cite{Husny06} has recently gained attention. The formation and the pinch-off mechanism of viscoelastic droplets in Newtonian continuous phases was investigated in various geometries of the flow-focusing device by Steinhaus {\it et al.}~\cite{Steinhaus}. The effect of the molecular weight of the polymer in the droplet phase on filament thinning was studied by Arratia {\it et al.}~\cite{Arratia08,Arratia09}. Typically, it is observed that elasticity of the droplet liquid prolongs the processes of thinning of the filament and significantly increases the interval required for breakup to complete. In a recent paper Garstecki {\it et al.}~\cite{Garstecki13} presented an experimental comparative study of the effect of elasticity of the continuous liquid in generation of droplets in microfluidic flow-focusing devices. The authors find that the elasticity of the focusing liquid stabilizes the jets facilitating formation of smaller droplets, and leads to transitions between various regimes at lower ratios of flow and at lower values of the Capillary numbers in comparison to the Newtonian focusing liquids. Complementing these results with systematic investigations by varying deformation rates and fluid constitutive parameters would be of extreme interest. This is witnessed by the various papers in the literature~\cite{Demenech07,LiuZhang09,LiuZhang11} addressing these kind of problems with the help of numerical simulations. In a recent paper, Liu \& Zhang~\cite{LiuZhang09,LiuZhang11} performed lattice Boltzmann simulations of three-dimensional microfluidic cross-junctions at low Capillary numbers. A regime map was created to describe the transition from droplets formation at the cross-junction (DCJ), downstream of the cross-junction (DC), to stable parallel flows (PF). The influence of flow rate ratio, Capillary number, and channel geometry was then systematically studied in the squeezing-pressure-dominated DCJ regime. In the present paper we study the impact of NN continuous phases in the transition from DCJ to DC regime.

\section{Theoretical Model}
Our numerical approach is based on a hybrid combination of LBM and finite difference schemes, the former used to model two immiscible fluids with a variable viscous ratio, and the latter used to model viscoelasticity using the FENE-P constitutive equations~\cite{bird,Herrchen}. LBM have already been used to model droplet deformation problems~\cite{Xi99,vandersman08,Komrakova13,Liuetal12}, droplets in confined T-junctions~\cite{Liuetal12,Gupta10}  and also viscoelastic flows~\cite{Onishi2,Malaspinas10}. This approach has been studied and validated in other dedicated works~\cite{SbragagliaGuptaScagliarini,SbragagliaGupta}, where we have provided evidence that the model is able to capture quantitatively rheological properties of dilute suspensions as well as deformation and orientation of single droplets in confined shear flows. We just recall here the main reference equations which are integrated in both the continuous and dispersed phases.\\  
In the continuous phase we integrate both the NS (Navier-Stokes) equations for the velocity ${\bm u_{c}}$ and the FENE-P equations 
\begin{eqnarray}\label{EQB}
\rho_c \left[ \partial_t \bm u_{c} + ({\bm u}_{c} \cdot {\bm \nabla}) \bm u_{c} \right] 
&=&  - {\bm \nabla}P_{c}+ {\bm \nabla} \left(\eta_{c} ({\bm \nabla} {\bm u}_{c}+({\bm \nabla} {\bm u}_{c})^{T})\right)+
              \frac{\eta_{P}}{\tau_{P}}{\bm \nabla} \cdot [f(r_{P}){\bm {\bm {\mathcal C}}}];  
                                                 \label{NSb}\\
\partial_t {\bm {\mathcal C}} + (\bm u_{c} \cdot {\bm \nabla}) {\bm {\mathcal C}}
&=& {\bm {\mathcal C}} \cdot ({\bm \nabla} {\bm u}_{c}) + 
                {({\bm \nabla} {\bm u}_{c})^{T}} \cdot {\bm {\mathcal C}} - 
                \frac{{f(r_{P}){\bm {\mathcal C}} }- {{\bm I}}}{\tau_{P}}.
                                                   \label{FENEb}
\end{eqnarray}
Here, $\eta_c$ is the dynamic viscosity of the continuous solvent phase, $\eta_{P}$ the viscosity parameter for the FENE-P solute, $\tau_P$ the polymer relaxation time, $\rho_c$ the solvent density, $P_c$ the solvent bulk pressure, $({\bm \nabla} {\bm u}_c)^T$ the transpose of $({\bm \nabla} {\bm u}_c)$, ${\bm {\mathcal C}}$ the polymer-conformation tensor, ${\bm I}$ the identity tensor, $f(r_P)\equiv{(L^2 -3)/(L^2 - r_P^2)}$ the FENE-P potential that ensures finite extensibility, $ r_P \equiv \sqrt{Tr({\bm {\mathcal C}})}$ and $L$ is the maximum possible extension of the polymers~\cite{bird}. In the dispersed phase we just consider the NS equations
\begin{eqnarray}\label{EQ}
\rho_d \left[ \partial_t \bm u_{d} + ({\bm u}_{d} \cdot {\bm \nabla}) \bm u_{d} \right] 
&=&  - {\bm \nabla}P_{d} + {\bm \nabla} \left(\eta_{d} ({\bm \nabla} {\bm u}_{d}+({\bm \nabla} {\bm u}_{d})^{T})\right)                                              
\end{eqnarray}
with $\eta_{d}$ the shear viscosity of the dispersed phase. Immiscibility between the droplet phase and the matrix phase is introduced using the so-called ``Shan-Chen'' model~\cite{SC93,SC94,SbragagliaGuptaScagliarini} which ensures phase separation with the formation of stable interfaces between the two phases characterized by a positive surface tension $\sigma$. As for the geometry used, we consider the simplest case where channels have a square cross-section $H$.  The square cross-section is resolved with $H \times H = 50 \times 50$ grid points, while the main channels and the side channel are resolved with $1150$ and $250$ grid points, respectively. We define, the flow ratio as $Q=\frac{v_d}{v_c} = \frac{Q_d}{Q_c}$, the  Capillary number as $Ca = \frac{\eta_c v_c}{\sigma}$, Reynolds number as $Re =\rho_c v_c H/ \eta_c$, where $Q_d=v_d H^2$ and $Q_c =v_c H^2$ are the flow rates at the two inlets;  $v_d$ and $v_c$ are the mean speeds in the inlets of the main and side channels, respectively, and $\sigma$ is the interfacial tension. In all the numerical simulations, the viscous ratio between the disperse phase ($\eta_d$) and the matrix phase ($\eta_c+\eta_P$) is set to $\lambda={\eta_d/(\eta_c+\eta_P)} = 1$, with $\eta_P/\eta_d = 0.4$.
The degree of viscoelasticity is computed from the Deborah number $\mbox{De}=\frac{N_1 H}{2 \sigma}\frac{1}{\mbox{\mbox{Ca}}^2}$, where  $N_1$ is the first normal stress difference which develops in homogeneous steady shear~\cite{bird,Lindner03}. In the Oldroyd-B limit ($L^2 \gg 1$), the Deborah number becomes $De=\frac{\tau_P}{\tau_{\mbox{\tiny{em}}}} \frac{\eta_P}{\eta_{d}}$, which is clearly dependent on the ratio between the polymer relaxation time $\tau_P$ and the emulsion time $\tau_{\mbox{\tiny{em}}}=\frac{H (\eta_{c}+\eta_P)}{\sigma}=\frac{H \eta_{d}}{\sigma}$.  All the simulations described in this paper refer to a case with $L^2=10^4$ (in LBM units). A proper tuning of the density gradients in contact with the wall allows for the modelling of the wetting properties which are chosen in such a way that the continuous phase completely wets the wall surfaces while the dispersed phase is non-wetting.


\section{Results and Discussions}

\begin{figure}[!t]
\makeatletter
\def\@captype{figure}
\makeatother
\subfigure[\,\,$t=t_0+150 \tau_{\mbox{\tiny{em}}}$, $Q = 2.5$]
{
\includegraphics[height=.09\textheight]{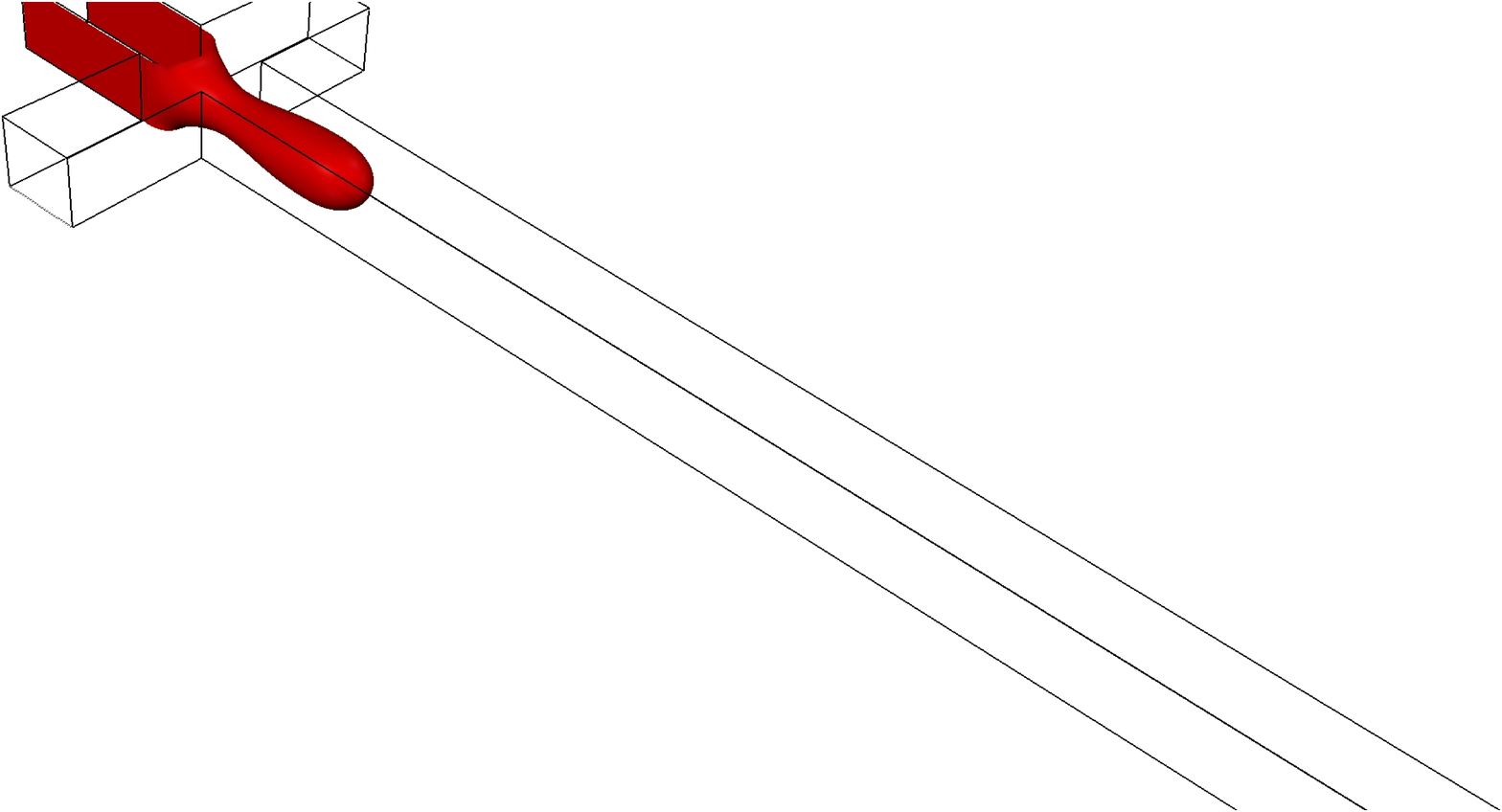}
}
\subfigure[\,\,$t=t_0+240 \tau_{\mbox{\tiny{em}}}$, $Q = 2.5$]
{
\includegraphics[height=.09\textheight]{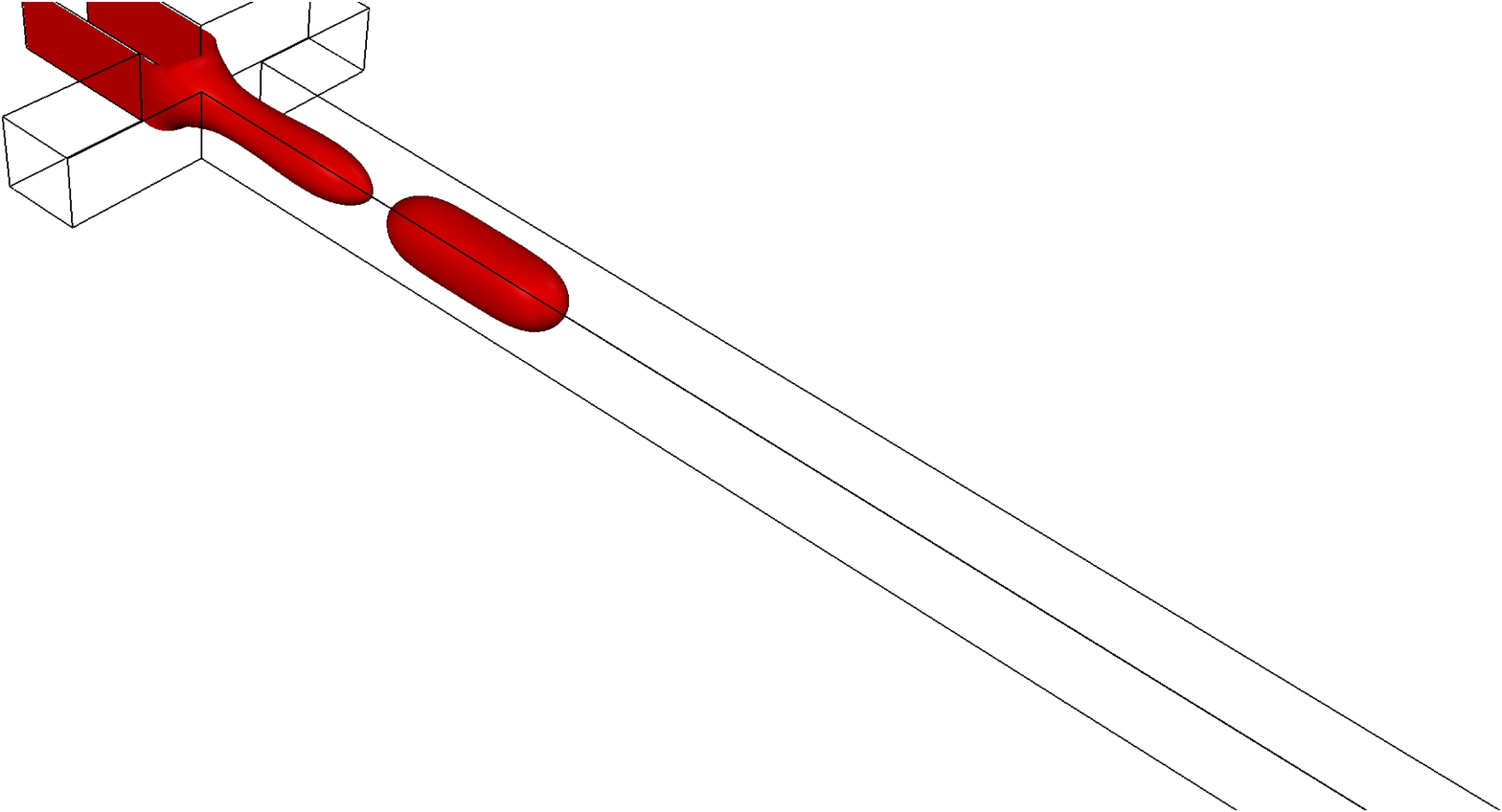}
}
\subfigure[\,\,$t=t_0+410 \tau_{\mbox{\tiny{em}}}$, $Q = 2.5$]
{
\includegraphics[height=.09\textheight]{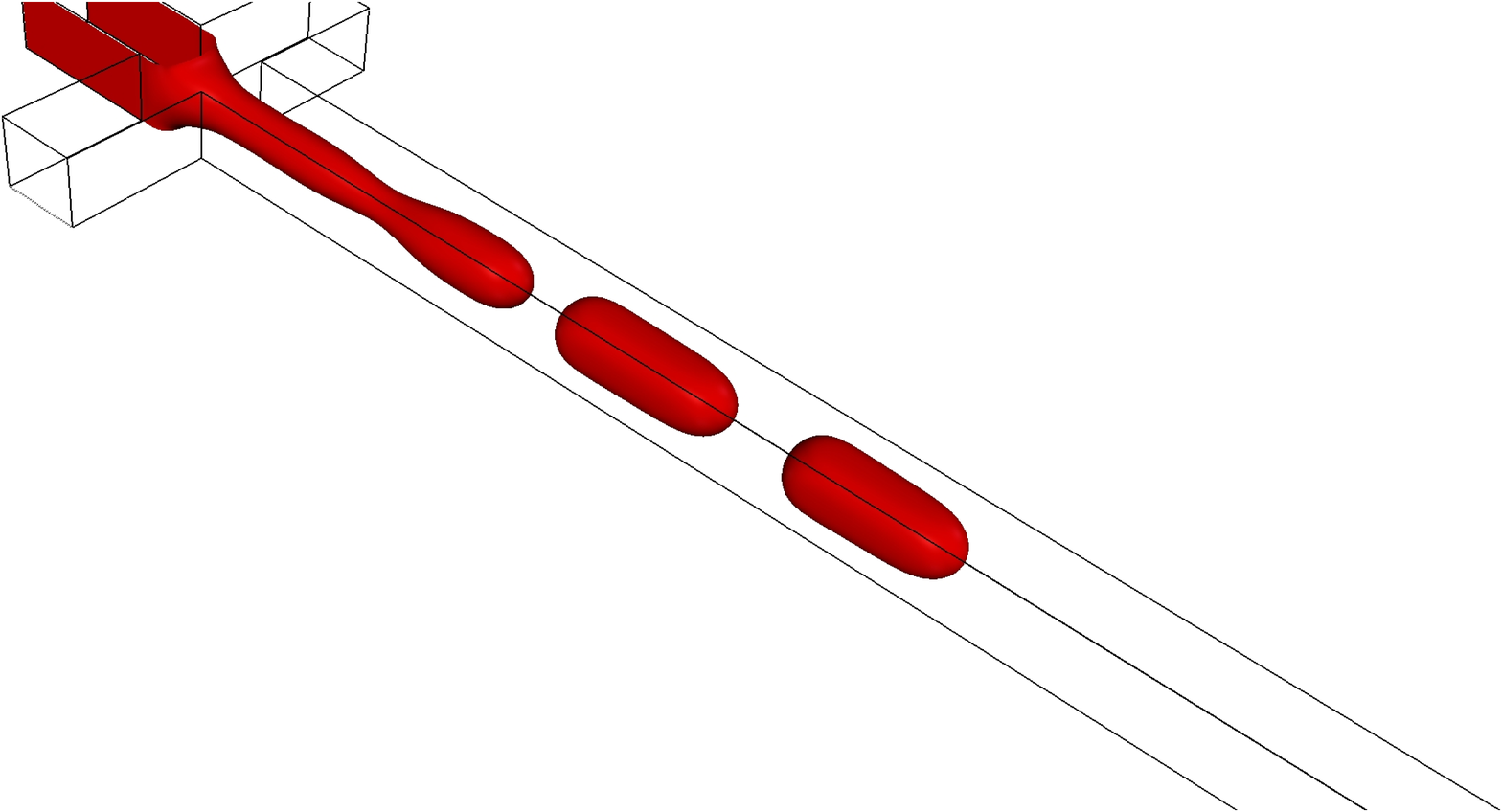}
}
\subfigure[\,\,$t=t_0+550 \tau_{\mbox{\tiny{em}}}$, $Q = 2.5$]
{
\includegraphics[height=.09\textheight]{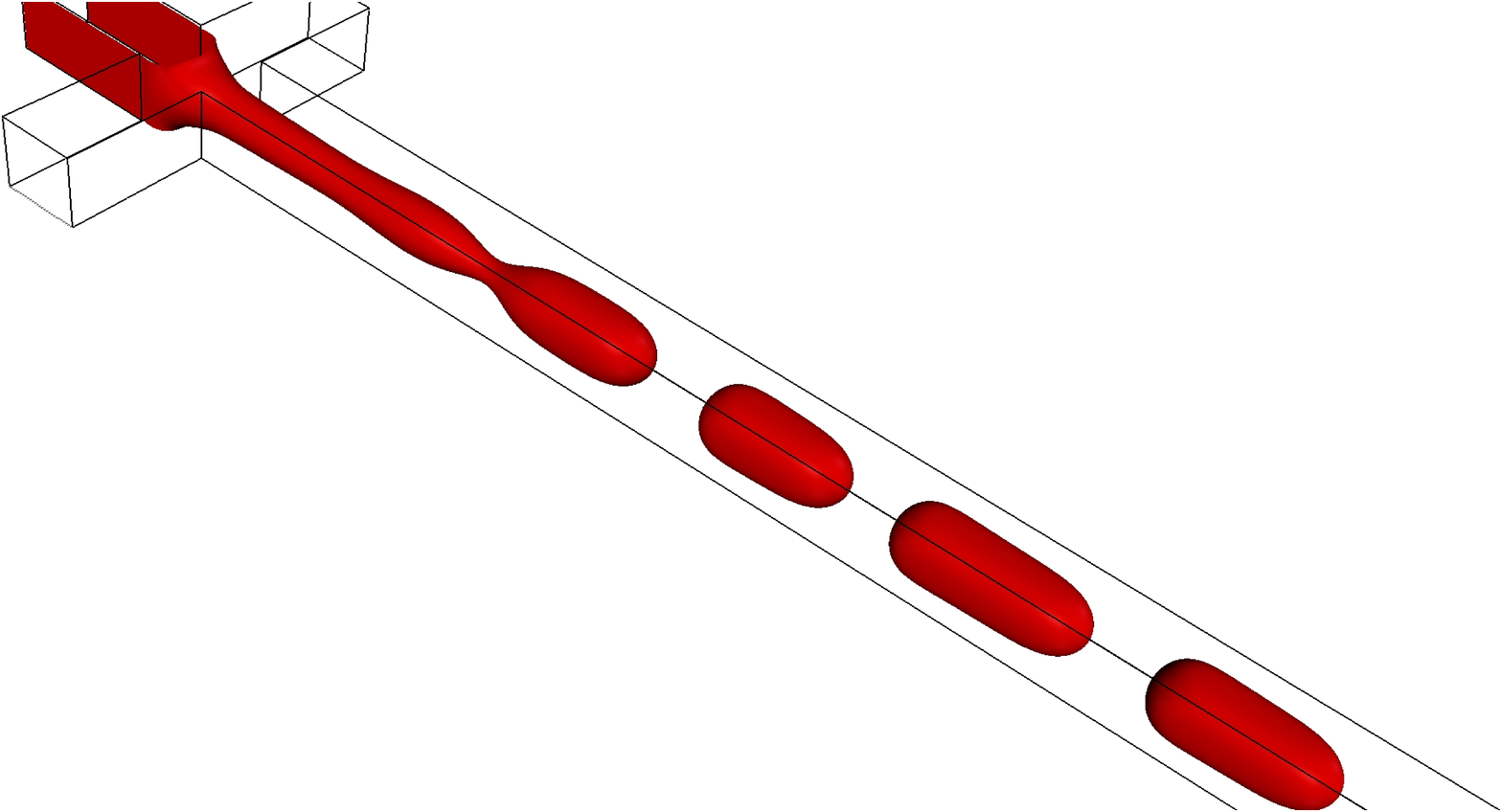}
}
\caption{Dynamics and Break-up of liquid threads in flow-focusing geometry from LBM simulations. We report the time history for 4 representative time frames (a) prior to the first break-up, (b) in post-break-up conditions, (c) the droplet after the second break-up, and (d) finally the droplet prior to the fourth break-up. The case reported in this figure corresponds to Newtonian disperse phase ($\eta_P=0$), $\mbox{Ca}=0.0056$, $Re=0.015$, $\lambda=1$ and $Q=2.5$. The various snapshots show the time history of the liquid thread dynamics and break-up. We have chosen the emulsion time as unit of time (see text for details).}
\label{fig:2}
\end{figure}

In this section we report the results for the break-up of threads in a confined flow-focusing device at fixed Capillary number. In a recent paper, Liu \& Zhang~\cite{LiuZhang09,LiuZhang11} performed LBM simulations and systematically characterized three typical flow patterns for different flow rate ratios at a fixed capillary number. At very low flow rate ratio $Q$, the droplets are formed at the cross-junction (DCJ) due to the squeezing mechanism~\cite{Garstecki06}. Upon increasing  $Q$, droplets are found to pinch-off downstream at the cross-junction (DC), forming a thread that becomes unstable after a distance $L_b$ from the cross-junction itself. Operatively, $L_b$ is computed as the distance between the break-up point and the corner up-stream at the cross-junction. As the flow rate ratio $Q$ increases to a critical value, stable parallel flows (PF) are observed, where the three incoming streams co-flow in parallel to the downstream without pinching. In addition, the transitions from DCJ to DC and from DC to PF are influenced by the Capillary number. As $Ca$ increases, the threshold value of flow rate ratio at which the transition occurs decreases, and the width of the DC regime also decreases. Droplet formation and break-up in the DC regime is shown in Fig. \ref{fig:2} for $Ca = 0.0056$, $Q=2.5$ for the case of a Newtonian Droplet in Newtonian matrix ($\eta_P=0$ in Eq. \eqref{FENEb}). Panels (a)-(d) show the droplet deformation prior to the first break-up at time $t = t_0 + 150 \tau_{\mbox{\tiny{em}}}$,  the droplet in post-break-up conditions at $t = t_0 + 240 \tau_{\mbox{\tiny{em}}}$,  the droplet after the second break-up at $t = t_0 + 410 \tau_{\mbox{\tiny{em}}}$,  and finally the droplet prior to the fourth break-up at $t = t_0 + 550 \tau_{\mbox{\tiny{em}}}$, respectively. As we can see from the Fig.~\ref{fig:2} the dispersed thread actually becomes unstable after a distance $L_b$ from the cross-junction and this distance increases as a function of time, which is another signature of the DC regime.


\begin{figure}[h!]
\makeatletter
\def\@captype{figure}
\makeatother
\subfigure[\,\,$t=t_0+550 \tau_{\mbox{\tiny{em}}}$, $De = 0$]
{
\includegraphics[height=.12\textheight]{newt_550000.eps}
}
\subfigure[\,\,$t=t_0+550 \tau_{\mbox{\tiny{em}}}$, $De = 0.4$]
{
\includegraphics[height=.12\textheight]{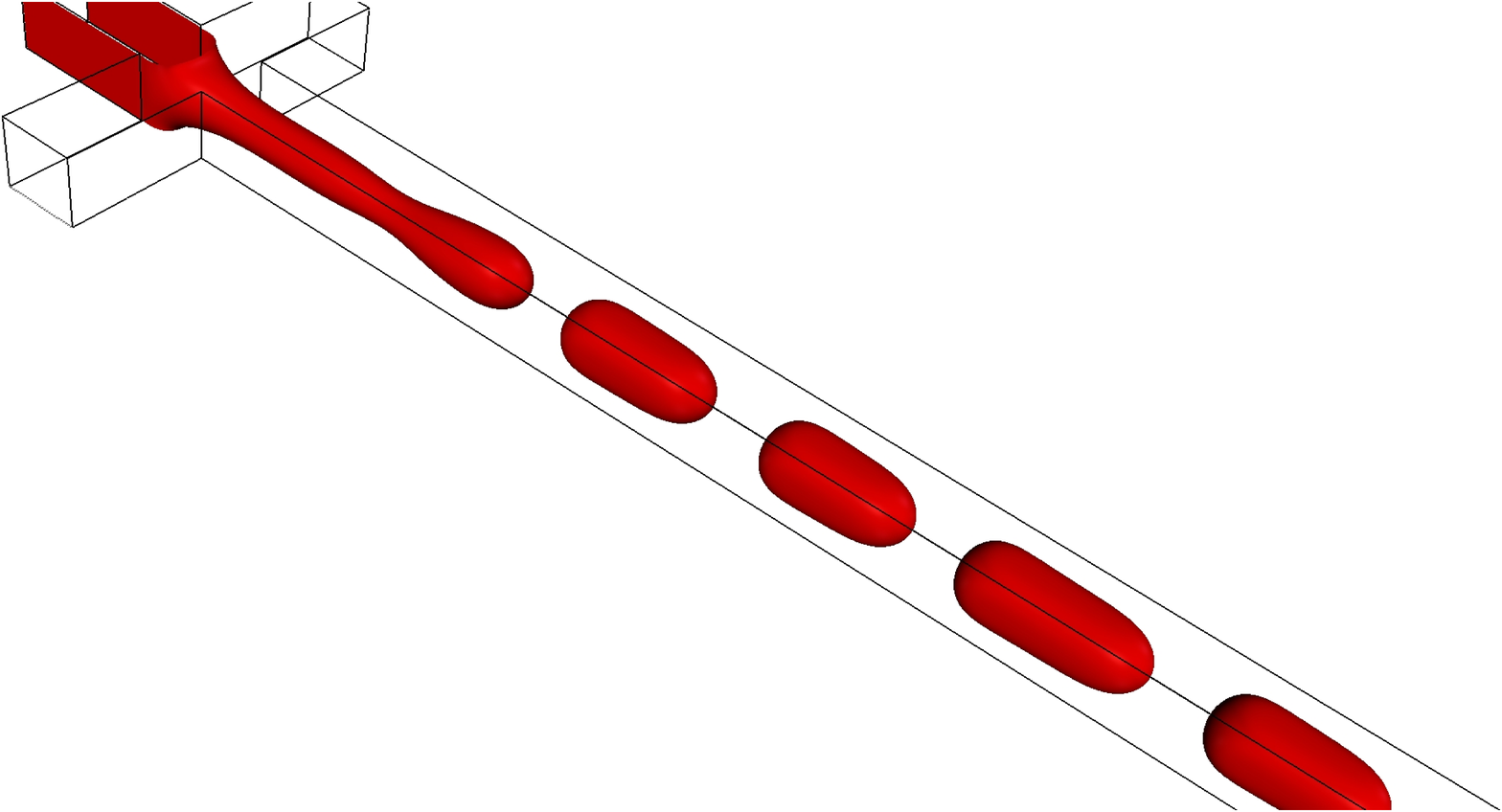}
}
\subfigure[\,\,$t=t_0+550 \tau_{\mbox{\tiny{em}}}$, $De = 2.0$]
{
  \includegraphics[height=.12\textheight]{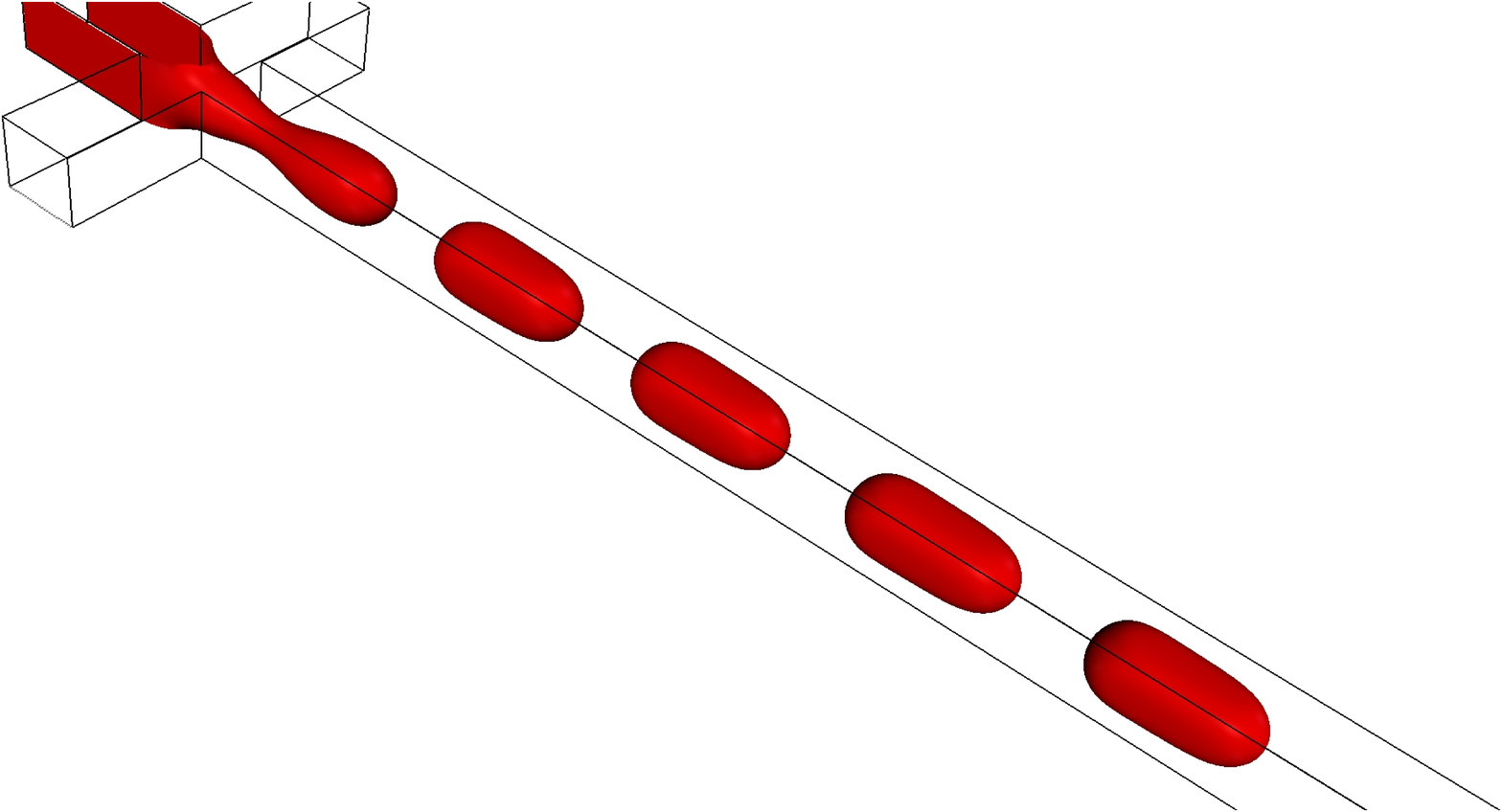}
}
\caption{Droplet formation in flow-focusing geometry at fixed $\mbox{Ca}=0.0056$, $Re=0.015$, $\lambda=1$ and $Q=2.5$. Three cases are compared: (a) Newtonian case ($De = 0.0$) at time $t = t_0 + 550 \tau_{\mbox{\tiny{em}}}$; (b) slightly viscoelastic case with $De = 0.4$ at time $t = t_0 + 550 \tau_{\mbox{\tiny{em}}}$; (c) viscoelastic case with $De = 2.0$ at time $t = t_0 + 550 \tau_{\mbox{\tiny{em}}}$.
} \label{fig:3}
\end{figure}

To appreciate the effects of viscoelasticity, in Fig.~\ref{fig:3} we show and compare the break-up process for different $De$. We report the density contours of the dispersed phase for three cases: (a) Newtonian matrix  at time $t=t_0+550 \tau_{\mbox{\tiny{em}}}$; (b) slightly viscoelastic matrix with $De = 0.4$ at time $t=t_0+550 \tau_{\mbox{\tiny{em}}}$;  (c) viscoelastic matrix with Deborah number just above unity ($De = 2.0$) at time $t=t_0+550 \tau_{\mbox{\tiny{em}}}$. All the other flow parameters are kept fixed, $\mbox{Ca}=0.0056$, $Re=0.015$, $\lambda=1$ and $Q=2.5$. We observe that as soon as the degree of viscoelasticity sensibly increases ($\mbox{De}=2.0$), the breaking point of the liquid thread approaches the cross-junction (Panel (c)), whereas for $\mbox{De}=0.4$ we only slightly perceive the effects of viscoelasticity. By comparing Panels (a) and (c) of Fig. \ref{fig:3} we also acknowledge a decrease in the droplet size, although such effect is not remarkable. By further increasing $Q$, also for the case of $\mbox{De}=2.0$ the breaking point starts shifting downstream at the cross-junction. All these observations lend support to the fact that the effect of viscoelasticity results in a delayed transition from the DCJ regime to the DC regime.


\section{Conclusions}
Microfluidic technologies offer the possibility to generate small fluid volumes of dispersed phases (droplets) in continuous phases.  One of the most common droplet generator is the flow-focusing device, based on the focusing a stream of one liquid in another co-flowing immiscible liquid. The confinement that naturally accompanies these devices has an impact on droplet deformation and break-up, which are significantly different from those of unbounded droplets. Moreover, in real processing conditions, relevant constituents have commonly a viscoelastic -rather than Newtonian- nature. In this paper we have presented results based on numerical simulations with the "lattice Boltzmann models" (LBM) to highlight the non trivial role played by confinement and non-Newtonian effects. We have worked in the operating conditions previously studied by other authors~\cite{LiuZhang09,LiuZhang11,Guillot}, where Newtonian droplets in a Newtonian matrix are formed either at the cross-junction, due to the squeezing mechanism~\cite{Garstecki06}, or downstream at the cross-junction (DC), forming a thread that becomes unstable after a distance $L_b$ from the cross-junction.  The transition between these two regimes has been found to be affected by matrix viscoelasticity. Future investigations on the role of droplet viscoelasticity are definitively worth of being considered. We also remark that the results presented in this paper are obtained for a fixed $L^2$, i.e. for a fixed maximum elongation of the polymers. The parameter $L^2$ is actually related to the extensional viscosity of the polymers~\cite{bird,Lindner03}. In the spirit of the work that we recently developed for characterizing droplet deformation and break-up in confined shear flow~\cite{SbragagliaGupta}, it would be of extreme interest to study the impact of a change in the finite extensibility parameter $L^2$ for the geometries studied in this paper~\cite{Arratia08,Arratia09}. Finally, we wish to underscore the role played by numerical simulations: in real experiments all the various processes (hydrodynamic forces, viscoelastic effects, confinement effects) occur at the same time and it is next to impossible to separately quantify their relative importance. The opportunity offered by numerical studies is to allow for a systematic analysis of each of the effects separately.

\begin{theacknowledgments}
We kindly acknowledge funding from the European Research Council under the Europeans Community's Seventh Framework Programme (FP7/2007-2013) / ERC Grant Agreement  N. 279004
\end{theacknowledgments}



\bibliographystyle{aipproc}   

\bibliography{sample}


\end{document}